\documentclass[letterpaper,twocolumn,10pt]{article}
\usepackage{usenix-2020-09}
\usepackage[available,functional,reproduced]{usenixbadges}

\usepackage{graphicx}
\usepackage{booktabs}
\usepackage{listings}
\usepackage{enumitem}
\usepackage{subcaption}
\IfFileExists{xurl.sty}{\usepackage{xurl}}{%
  \PassOptionsToPackage{hyphens}{url}\usepackage{url}%
  \makeatletter
  \g@addto@macro\UrlBreaks{\do\-\do\.\do\=\do\?\do\&\do\_\do\:\do\;%
    \do\,\do\!\do\'\do\(\do\)\do\*\do\+\do\$\do\#\do\@\do\%\do\~%
    \do\/\do\a\do\b\do\c\do\d\do\e\do\f\do\g\do\h\do\i\do\j\do\k%
    \do\l\do\m\do\n\do\o\do\p\do\q\do\r\do\s\do\t\do\u\do\v\do\w%
    \do\x\do\y\do\z\do\A\do\B\do\C\do\D\do\E\do\F\do\G\do\H\do\I%
    \do\J\do\K\do\L\do\M\do\N\do\O\do\P\do\Q\do\R\do\S\do\T\do\U%
    \do\V\do\W\do\X\do\Y\do\Z\do\0\do\1\do\2\do\3\do\4\do\5\do\6%
    \do\7\do\8\do\9}%
  \makeatother
  \setlength{\emergencystretch}{3em}}
\IfFileExists{pgf-umlsd.sty}{\usepackage{pgf-umlsd}}{}

\usepackage{amsmath}
\usepackage{multirow}
\usepackage{tabularx}
\usepackage{tikz}
\usetikzlibrary{arrows,arrows.meta, positioning, shapes.geometric, fit}

\IfFileExists{todonotes.sty}{\usepackage{todonotes}}{%
  \providecommand{\todo}[2][]{}\providecommand{\missingfigure}[1]{}}
\IfFileExists{url.sty}{\usepackage{url}}{\providecommand{\path}[1]{\texttt{#1}}}

\newcommand{\macOS}{\mbox{macOS}}
\newcommand{\iOS}{\mbox{iOS}}
\newcommand{\WiFi}{\mbox{Wi-Fi}}

\begin{document}

\date{}

% Double-blind: no author names
\title{\Large \bf Protocol Prying: Systematic Vulnerability Research in the Apple AirDrop\\and Android Quick Share Proximity Transfer Protocols}

\author{
  Arash Ale Ebrahim \qquad\qquad Nils Ole Tippenhauer  \\[0.6em]
  CISPA Helmholtz Center for Information Security
}

\maketitle
\thispagestyle{empty}

% ============================================================================
% ABSTRACT
% ============================================================================
\begin{abstract}
Apple AirDrop and Google/Samsung Quick Share are proximity file-transfer protocols used by over five billion devices, yet their application-layer security properties remain largely unstudied because both stacks are proprietary and undocumented.
Both protocols are reachable from wireless proximity without any prior pairing and process complex serialized content (binary plists, CPIO archives, Protocol Buffers, UKEY2 handshakes) inside privileged daemons, making them attractive zero-click targets across multiple operating systems.
We perform the first cross-platform reverse engineering and protocol-aware fuzzing study of both stacks. We reconstruct AirDrop's seven-layer state machine and DVZip adaptive compression from binary analysis, build \textsc{AirFuzz}, a protocol-aware fuzzer that mutates pre-compression representations, and complement it with targeted hand-written analyses of Samsung's Quick Share service and Google's Quick Share for Windows.
We discover six vulnerabilities (V1-V6): three pre-authentication issues in \macOS{}/\iOS{} AirDrop (V1: Swift \texttt{fatalError} DoS in the HTTP path router; V2: unbounded XML plist recursion in \texttt{Foundation}, V3: NULL deref in \texttt{Network.framework}'s HTTP/1.1 parser), two protocol-layer flaws in Samsung Quick Share (V4: pre-authentication \texttt{OfflineFrame} dispatch, V5: D2D encryption bypass for three frame types), and a heap use-after-free in Google Quick Share for Windows (V6) for which Google awarded a bounty. We responsibly disclosed all findings, Apple, Samsung, and Google have acknowledged the reports.
For transparency and reproducibility, the AirFuzz artifacts are publicly available on Zenodo: \url{https://zenodo.org/records/20442029}.
\end{abstract}

\section{Introduction}
\label{sec:introduction}

Proximity-based file transfer is a fundamental feature of modern mobile and desktop operating systems. Apple's AirDrop, introduced in 2011, and Google's Nearby Share (now Quick Share), launched in 2020, collectively serve over five billion active devices~\cite{apple-active-devices, google-android-devices}. These protocols enable ad hoc peer-to-peer file transfer between nearby devices over a combination of Bluetooth Low Energy (BLE) for discovery and Wi-Fi for data transfer, without requiring shared network infrastructure.

From a security perspective, proximity transfer protocols present a compelling attack surface~\cite{beer2020awdl,silvanovich2019zeroclick,gross2020remote,yair2024quickshell,toothpicker}. The attacker need only be within wireless range (typically 10--30\,m for AWDL/\WiFi{} Direct), and the initial protocol phases (service discovery and connection establishment) require \emph{no authentication}. On Apple devices configured with ``Everyone'' visibility, the entire Discover and Ask phases are reachable without any user interaction or prior pairing. This creates a zero-click, pre-authentication attack surface that extends from the privileged \texttt{sharingd} daemon (which also manages AirPlay, Handoff, Universal Clipboard, and Continuity Camera) all the way down into the kernel: Beer's iOS zero-click radio-proximity exploit~\cite{beer2020awdl} demonstrated remote code execution against the AWDL kernel driver itself, showing that the attack surface is not limited to user-space sharing code.

Despite the ubiquity and security relevance of these protocols, prior academic work has focused primarily on privacy aspects of AirDrop's contact discovery mechanism~\cite{heinrich2021opendrop, heinrich2021privatedrop, stute2021disrupting, martin2019handoff} and the underlying AWDL link layer. The foundational work of Stute~et~al.~\cite{stute2018billion,stute2019openwifi,stute2021disrupting} reverse engineered AWDL itself, released the open-source OWL stack, and demonstrated link-layer denial-of-service and man-in-the-middle attacks; their results provide the wireless substrate on which our application-layer analysis builds. In parallel, Project~Zero~\cite{beer2020awdl} weaponised the AWDL kernel attack surface against \iOS{}. A comprehensive security analysis of the \emph{application-layer} AirDrop protocol (binary property lists, DVZip compression, CPIO archive extraction, and the HTTP-layer state machine in \texttt{sharingd}) has not been published. Similarly, while Google's Quick Share has received attention following the SafeBreach RCE chain in its Windows client~\cite{yair2024quickshell}, the Samsung Android implementation's security properties remain largely unexamined.

In this paper, we bridge this gap with a systematic security analysis of both Apple AirDrop and Android Quick Share. Our work makes the following contributions:

\begin{enumerate}[leftmargin=*,topsep=2pt,itemsep=1pt]
    \item \textbf{Application-layer reverse engineering of AirDrop.} Building on the AWDL link-layer work of Stute~et~al.~\cite{stute2018billion,stute2019openwifi}, we reverse engineer the complete \emph{application-layer} AirDrop stack on current \macOS{} and \iOS{}, documenting seven protocol layers from AWDL up through DVZip/CPIO archive extraction. We reconstruct the receiver-side state machine of \texttt{sharingd}, document the previously unpublished DVZip adaptive chunked compression format, and catalogue 40+ protocol commands. Detailed offsets, disassembly, and symbol lists for specific binary versions are released as artifacts rather than reproduced inline.

    \item \textbf{Protocol-Aware Fuzzer.} We develop \textsc{AirFuzz}, a protocol-aware fuzzer with nine mutation layers (HTTP, binary plist, DVZip, CPIO, DER, havoc, memcorrupt, and field-level) and Frida-based coverage collection. We discuss why we did not use Frida \texttt{Stalker}, \textsc{fpicker}~\cite{fpicker,toothpicker}, or \textsc{TinyInst}/\textsc{Jackalope}~\cite{tinyinst,jackalope,p0-coreaudiod} as our coverage engine; in particular, we reproduce the Stalker breakage on arm64e \macOS{}/\iOS{} (Section~\ref{sec:fuzzing}).

    \item \textbf{AirDrop Vulnerabilities.} We discover two zero-click, and one post-accept pre-authentication vulnerabilities: (V1)~a remote DoS via Swift \texttt{fatalError} in the HTTP path router, (V2)~a stack overflow in \texttt{Foundation}'s XML property list parser and (V3)~a NULL pointer dereference in \texttt{Network.framework}.

    \item \textbf{Quick Share Vulnerabilities.} We discover three vulnerabilities in Quick Share: (V4)~a pre-authentication frame processing bypass that allows unauthenticated protocol interaction before UKEY2 handshake completion, (V5)~a device-to-device (D2D) encryption bypass where 3 out of 7 post-handshake frame types are processed without the mandatory \texttt{SecureMessage} encryption wrapper in Samsung's Android implementation,  and (V6)~a critical use-after-free via an endpoint collision race condition in Google Quick Share for Windows which leads to remote-code-execution (version~1.2.2472.1).

    \item \textbf{Cross-Platform Comparison.} We provide the first cross-platform security comparison of Apple and Google/Samsung proximity transfer implementations, identifying shared vulnerability patterns and divergent design decisions across six total vulnerabilities spanning macOS, iOS, Android, and Windows.
\end{enumerate}

We responsibly disclosed all vulnerabilities to Apple Product Security, Samsung Mobile Security, and the Google Vulnerability Reward Program. Apple acknowledged V1--V3 and remediation is in progress; Samsung transferred V4--V5 to Google; Google acknowledged V6 and awarded a bounty for the Windows Quick Share use-after-free, with a CVE pending assignment. Section~\ref{sec:ethics} provides the full disclosure timeline and vendor responses.

\paragraph{Vulnerability overview.}
Table~\ref{tab:vuln-overview} summarises the six vulnerabilities, their target component, the precondition required to reach them, and the observed impact. We use this table as a roadmap throughout the paper.

\begin{table*}[t]
\centering
\caption{Overview of the six vulnerabilities (V1--V6). \textbf{Precond.}: ``Everyone'' = AirDrop in ``Everyone for 10\,minutes'' visibility; ``Visible'' = Quick Share visible to nearby devices; ``On-path'' = same-LAN attacker (V5 only); ``Race'' = concurrent connect/disconnect (V6). \textbf{Impact}: DoS = denial of service; PoP = protocol-state manipulation; UAF = potentially exploitable use-after-free.}
\label{tab:vuln-overview}
\footnotesize
\setlength{\tabcolsep}{6pt}
\begin{tabular*}{\textwidth}{@{\extracolsep{\fill}}llllll@{}}
\toprule
\textbf{ID} & \textbf{Target} & \textbf{Component} & \textbf{Class} & \textbf{Precondition} & \textbf{Impact} \\
\midrule
V1 & AirDrop & \texttt{Sharing.framework} path router & Fatal assert & Everyone & DoS \\
V2 & AirDrop & \texttt{Foundation} XML plist & Stack overflow & Everyone & DoS \\
V3 & AirDrop & \texttt{Network.framework} HTTP/1.1 & NULL pointer dereference & Everyone & DoS \\
V4 & QuickShare & Samsung GMS dispatcher & Authentication bypass & Visible & PoP \\
V5 & QuickShare & Samsung D2D layer & Crypto bypass & On-path & PoP \\
V6 & QuickShare & Win.\ endpoint mgmt & UAF race & Race & RCE \\
\bottomrule
\end{tabular*}
\end{table*}

\section{Background}
\label{sec:background}

\subsection{Apple AirDrop}
\label{sec:bg-airdrop}

AirDrop, introduced in OS X Lion (2011) and iOS~7 (2013), enables peer-to-peer file transfer between Apple devices. It relies on two wireless technologies: \emph{Bluetooth Low Energy} (BLE) for initial device discovery and wake-up, and \emph{Apple Wireless Direct Link} (AWDL) for the actual data transfer. AWDL is a proprietary Wi-Fi-based protocol that creates an ad-hoc mesh network on the \texttt{awdl0} virtual interface using IPv6 link-local addresses~\cite{stute2018billion}.

The AirDrop application logic is implemented in \texttt{sharingd}, a privileged daemon that manages multiple Apple continuity services. On the receiver side, \texttt{sharingd} listens on TCP port~8770 over AWDL with TLS, serving an HTTP/1.1 API. The daemon is managed by \texttt{launchd} and automatically restarts on crash, but with exponentially increasing throttle delays after consecutive failures.

\texttt{sharingd} is a large, monolithic system daemon. On iOS~18.1/26.3 (ARM64e), The binary has \emph{stripped} only one exported symbols (\texttt{\_main} and \texttt{\_\_mh\_execute\_header}) are visible via standard tools. It dynamically loads the private \texttt{Sharing.framework} (3.8\,MB, 5,032 exported symbols), which implements the core AirDrop protocol logic, and links against approximately 30 system frameworks including \texttt{Foundation}, \texttt{Security}, \texttt{Network}, \texttt{CoreFoundation}, and \texttt{libswift\_Concurrency}. The daemon manages not only AirDrop but also AirPlay, Handoff, Universal Clipboard, Continuity Camera, NameDrop, and SharePlay, meaning a crash in any subsystem disrupts all Continuity services simultaneously.

AirDrop supports three visibility modes: \emph{Receiving Off} (no AWDL listener), \emph{Contacts Only} (requires Apple~ID certificate chain verification), and \emph{Everyone for 10 Minutes} (accepts connections from any device). In ``Everyone for 10 Minutes'' mode, the complete HTTP API is reachable without any form of authentication. On iOS, Apple added a 10-minute auto-timeout that reverts ``Everyone'' to ``Contacts Only'' , but this mitigation does not protect against attacks during the window of exposure, and does not apply to macOS.

\subsection{Android Quick Share}
\label{sec:bg-quickshare}

Quick Share (formerly Nearby Share) is Google's cross-platform file-sharing feature for Android, Chrome~OS, and Windows. On Samsung devices, it is deeply integrated as the default sharing mechanism, replacing Samsung's earlier ``Nearby Share'' implementation. Quick Share uses a combination of BLE, Wi-Fi Direct, Wi-Fi LAN, and may also use WebRTC depending on the scenario.

Quick Share’s session traffic is largely structured as Protobuf-serialized messages (e.g., the \texttt{OfflineFrame} family), and is typically protected by an authenticated key-exchange followed by encrypted transport. In particular, reverse-engineering of the Quick Share for Windows implementation reports an initial UKEY2 handshake after which subsequent \texttt{OfflineFrame} packets are encrypted. \cite{safebreachQuickShareRCE}

At a high level, the connection lifecycle can be described in four phases: (1)~\emph{Discovery / advertising}, where nearby devices become visible to each other using short-range radios (commonly Bluetooth/BLE) and exchange ephemeral endpoint identifiers; (2)~\emph{Bandwidth upgrade negotiation}, where the peers negotiate a higher-throughput out-of-band channel (e.g., Wi-Fi Direct or a temporary Wi-Fi hotspot / access point) for bulk transfer; (3)~\emph{Introduction \& consent}, where the sender transmits file-introduction metadata and the receiver is prompted to accept the transfer; and (4)~\emph{Payload transfer}, where the file bytes are transmitted over the negotiated high-bandwidth channel. \cite{yair2024quickshell,googleQuickShareHelp}

Samsung's implementation extends the base Quick Share with additional features including link sharing with server-side URL preview generation, a feature not present in Google's reference implementation. On the Samsung Galaxy~S23 Ultra (our test device), Quick Share runs as a system-level service (\texttt{com.samsung.android.nearby.discovery}) with access to network I/O, file system operations, and WebView rendering for link previews.

Google's Quick Share for Windows implementation handles endpoint management, the lifecycle of connected peers through a state machine that allocates and deallocates endpoint objects as devices connect and disconnect. During concurrent transfers involving multiple devices, these endpoint objects can be subject to race conditions if connection and disconnection events arrive simultaneously, a property we exploit in v6 (Section~\ref{sec:vuln-v6}).

\subsection{Threat Model}
\label{sec:threat-model}

We consider an attacker in wireless proximity to the target device (within AWDL/Wi-Fi Direct range, typically 10-30 meters). The attacker has a commodity laptop with Wi-Fi capability and requires no prior relationship with the target (no pairing, no contact exchange, no shared network). The target device has AirDrop or Quick Share enabled in a mode that accepts connections (``Everyone'' for AirDrop; default visibility for Quick Share). We focus on \emph{zero-click} attacks that require no user interaction on the target device. For AirDrop, this means exploiting the pre-authentication phases (Discover, Ask routing) before the user acceptance prompt is displayed.

\section{AirDrop Protocol Reverse Engineering}
\label{sec:protocol}

We reverse engineered the AirDrop protocol by static and dynamic analysis of \texttt{sharingd} and the private \texttt{Sharing.framework} on \macOS{}~15.7.4 and on the latest \macOS{}~26.3 and \iOS{}~18.1/26.3 releases. Throughout the paper we describe findings at the protocol level; exact symbol offsets and per-build disassembly listings are released as artifacts.

\subsection{Protocol Stack Overview}
\label{sec:protocol-stack}

Table~\ref{tab:protocol-stack} summarizes the seven-layer AirDrop protocol stack we identified. Each layer introduces distinct serialization formats and potential vulnerability surfaces.

\begin{table*}[t]
\centering
\caption{AirDrop protocol stack. Each layer adds serialization complexity and attack surface.}
\label{tab:protocol-stack}
\begin{tabular}{@{}clll@{}}
\toprule
\textbf{\#} & \textbf{Layer} & \textbf{Technology} & \textbf{Key Attack Surface} \\
\midrule
7 & Archive       & CPIO newc / BOM (Bill of Materials) & Path traversal, symbolic-link attacks \\
6 & Compression   & DVZip / gzip                        & Chunk-length validation, decompression bombs \\
5 & Encoding      & Binary property list (bplist)       & Type confusion, unbounded recursion \\
4 & Transport     & HTTP/1.1                            & Path routing, request smuggling \\
3 & Security      & TLS 1.2 / 1.3                       & Self-signed certificates, no client authentication \\
2 & Network       & IPv6 link-local                     & \texttt{fe80::} addresses on the \texttt{awdl0} interface \\
1 & Link          & AWDL / BLE                          & Wireless-proximity discovery \\
\bottomrule
\end{tabular}
\end{table*}

\subsection{Discovery and Connection Establishment}
\label{sec:discovery}

AirDrop device discovery uses DNS-SD (mDNS) over AWDL, advertising service type \texttt{\_airdrop.\_tcp.local.} with $\sim$25 TXT keys; the \texttt{flags} field is a bitmask (e.g., \texttt{0x3FB}) encoding DVZip, pipelined uploads, mixed content types, URL sharing, contact resolution, and BLE support. Upon discovery, the sender opens a TLS connection to port~8770 on the receiver's IPv6 link-local address over AWDL, using self-signed certificates with no hostname verification and no client certificate. \emph{Any device within AWDL range can establish a TLS connection and interact with the HTTP API} regardless of the receiver's visibility mode.

\subsection{HTTP API and State Machine}
\label{sec:state-machine}

The AirDrop HTTP API exposes seven POST endpoints:

\begin{center}
\small
\texttt{/Discover}, \texttt{/Hello}, \texttt{/Ask}, \texttt{/Upload},\\
\texttt{/Exchange}, \texttt{/SharedIdentity}, \texttt{/Error}
\end{center}

The receiver-side state machine is implemented in \texttt{SDAirDropConnection}. Each TLS connection has three \emph{request slots}---one each for \texttt{Discover}, \texttt{Ask}, and \texttt{Upload}---plus control flags and a \texttt{dispatch\_semaphore\_t} that blocks the request handler while the user-consent prompt is on screen. A slot is \texttt{NULL} until the corresponding request arrives. \emph{Discover} is accepted only when all three slots are \texttt{NULL} and returns a plist \texttt{StatusCode} of 100 (Everyone), 200 (contacts matched), or 401 (Contacts-only, no match). \emph{Ask} also requires all slots \texttt{NULL} (a prior Discover slot is cleared first), stores the payload, displays the consent prompt, and blocks on the semaphore until accept/decline or a 15\,s timeout. \emph{Upload} requires the Upload slot \texttt{NULL} and normally follows a successful Ask on the \emph{same} TLS connection; two TXT-record flags from discovery are consulted---\texttt{SupportsUPP} (bit \texttt{0xb}) permits Upload \emph{without} a prior Ask, and \texttt{SupportsDVZip} enables adaptive compression. Upload data \emph{streams into memory while the Ask prompt is displayed}; the semaphore gates processing, not reception.

\begin{figure}[t]
\centering
\includegraphics[width=0.8\columnwidth]{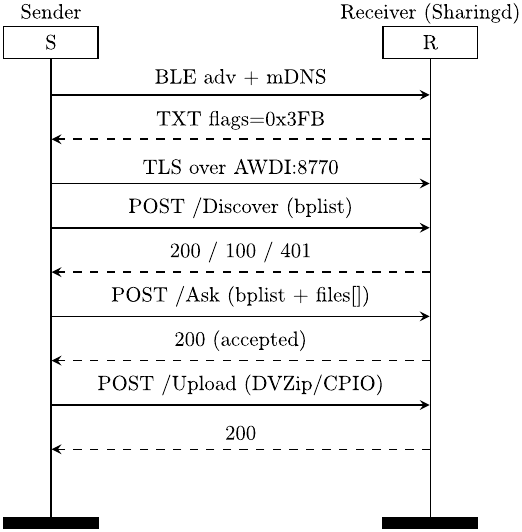}
\caption{End-to-end AirDrop exchange. Discover and Ask are reachable pre-consent on the same TLS connection. The receiver's \texttt{flags} TXT record (e.g., \texttt{0x3FB}, with bit \texttt{0xb}~=~\texttt{Supports\-UPP}) is consulted on Upload to decide whether pipelined uploads are allowed.}
\label{fig:airdrop-flow}
\end{figure}

\noindent A normal AirDrop transfer is therefore: Discover (optional) $\rightarrow$ Ask (same TLS connection) $\rightarrow$ Upload (same TLS connection). The protocol requires Ask and Upload to share the same TLS session; Upload on a separate connection is rejected. Figure~\ref{fig:airdrop-flow} shows this exchange end-to-end.

\subsubsection{Path Routing Vulnerability Surface}
\label{sec:path-routing}

The HTTP path router lives in \texttt{Sharing.framework} (file \path{Network+SFAirDropMessage.swift} in Apple's source layout) and dispatches incoming request paths via a Swift \texttt{switch} statement over an \texttt{SFAirDropMessagePath} enum. The \texttt{default} case calls \texttt{fatalError("Unhandled Path \textbackslash(path)")}, which the Swift compiler lowers to a trapping instruction that aborts the process on any architecture. Any unauthenticated client sending an HTTP request to an unrecognised path triggers this trap; we exploit it as V1 (Section~\ref{sec:vuln-v1}).

\subsection{Serialization Formats}
\label{sec:serialization}

\paragraph{Binary Property Lists.}
Discover and Ask request/response bodies are serialized as Apple binary property lists (\texttt{bplist00}), parsed by \texttt{CFPropertyListCreateWithData} and type-checked as \texttt{CFDictionary}.

\paragraph{DVZip Adaptive Compression.}
Upload bodies use DVZip (\texttt{Content-Type: application/x-dvzip}), an Apple-proprietary adaptive chunked format we identified through binary analysis: 4-byte big-endian chunk-length headers wrap individual zlib (deflate) blocks, with adaptive switching between gzip and DVZip mid-stream depending on compressibility, and graceful fallback to plain gzip when the receiver does not advertise \texttt{SupportsDVZip}.

\paragraph{CPIO Archives.}
The compressed payload contains a CPIO ``newc'' archive (magic \texttt{070701}) terminated by \texttt{TRAILER!!!}. After decompression, \texttt{sharingd} extracts files via Apple's BOM (Bill of Materials) library using \texttt{BOMCopierCopyWithOptions}, with CPIO filenames passed to \texttt{BOMFSObjectNewFromPath}---the surface for path-traversal, symlink, and filename-based attacks.

\subsection{Authentication and Identity}
\label{sec:auth}

\subsubsection{Protocol Keys (summary).}
Through string extraction and cross-referencing with disassembly we identified 40+ protocol commands in AirDrop's binary-property-list messages, grouped into Discover keys (e.g., \texttt{SenderComputerName}, \texttt{SenderModelName}, \texttt{SenderRecordData}---the DER-encoded Apple~ID certificate chain, \texttt{SenderID}, \texttt{TransferVersion}), Ask keys (\texttt{Files}, \texttt{FileType}, \texttt{FileName}, \texttt{FileBomPath}, \texttt{ConvertMediaFormats}), and capability flags (\texttt{SupportsDVZip}, \texttt{SupportsUPP}, \texttt{SupportsMixedTypes}, \texttt{SupportsURLSharing}, \texttt{SupportsStreamZip}). The complete key list with descriptions is released as an artifact. The three security-sensitive keys are \texttt{SenderRecordData} (parses untrusted DER), \texttt{FileBomPath} (path-traversal surface, mitigated by \texttt{BOMCopierCopyWithOptions} sandboxing), and \texttt{TransferVersion} (gates DVZip and pipelined-upload behaviour).

\subsubsection{AirDrop Authentication Model}

AirDrop's authentication is layered. \emph{TLS} uses self-signed certificates and provides encryption only, not peer verification. \emph{Contact resolution} parses the DER-encoded Apple~ID certificate chain in \texttt{SenderRecordData} via \texttt{SFAppleIDVerifyCertificateChain} to decide whether the sender is a known contact. \emph{Auto-accept} is gated by the \texttt{canAutoAccept} predicate, whose \texttt{senderIsMe} flag is set only when \texttt{SFAppleIDVerifyCertificateChain} succeeds against the local user's iCloud-bound certificate and the resulting Apple~ID matches the receiver's; \texttt{senderIsMe} is \emph{not} a plist field controlled by the sender. \emph{User consent} is enforced by the Ask-phase UI prompt gated on a \texttt{dispatch\_semaphore}; all file delivery requires either user acceptance or same-Apple-ID auto-accept. We tested 15+ bypass variants that manipulate plist-level fields purporting to set \texttt{senderIsMe} or replace \texttt{SenderRecordData}; all were rejected (Section~\ref{sec:protocol-zero-click}).

\subsubsection{Zero-Click Bypass Analysis}
\label{sec:protocol-zero-click}

We systematically tested ten strategies to bypass the user-acceptance requirement---direct Upload skipping Ask, HTTP-pipelined atomic Ask$+$Upload, plist-field manipulation of \texttt{SenderIsMe}, forged \texttt{SenderRecordData}, fifteen further Ask-field manipulations, notification flooding, repurposing the Exchange/NameDrop \texttt{TransferType}, Upload on a separate TLS session, chunked-Upload interleaving, and forged \texttt{TransferID}. \emph{All ten} were rejected or timed out on \iOS{}~18.1/26.3 and \macOS{}~15.7.4/26.3: the auto-accept decision is gated behind \texttt{SDAppleIDAuthenticateCertificateChainSync}, which verifies the sender's Apple~ID against a hardware-bound chain that cannot be forged without access to the sender's iCloud credentials. V1 (Section~\ref{sec:vuln-v1}) instead abuses the unrecognised-path trap before any state transition occurs.

\section{Fuzzing Methodology}
\label{sec:fuzzing}

\subsection{Challenges}
\label{sec:fuzz-challenges}

Fuzzing AirDrop presents four unique challenges compared to traditional network protocol fuzzing:

\begin{enumerate}[leftmargin=*,topsep=2pt,itemsep=1pt]
  \item \textbf{AWDL transport}: The target is only reachable over the \texttt{awdl0} interface via IPv6 link-local addresses, requiring active AWDL participation (BLE advertising + Wi-Fi channel synchronization).
  \item \textbf{TLS requirement}: All HTTP communication is over TLS with self-signed certificates, requiring a proper TLS handshake for each connection.
  \item \textbf{Multi-layer encoding}: A valid Upload requires DVZip-compressed CPIO archives containing binary plist metadata-mutations at any single layer are likely rejected by parsers at other layers.
  \item \textbf{State dependency}: Upload requires a prior Ask on the \emph{same} TLS connection, and the Ask phase blocks on user interaction (15-second timeout without our auto-accept hook).
\end{enumerate}

\subsection{\textsc{AirFuzz} Architecture}
\label{sec:airfuzz-arch}

\begin{figure*}[t]
\centering
\includegraphics[width=0.95\textwidth]{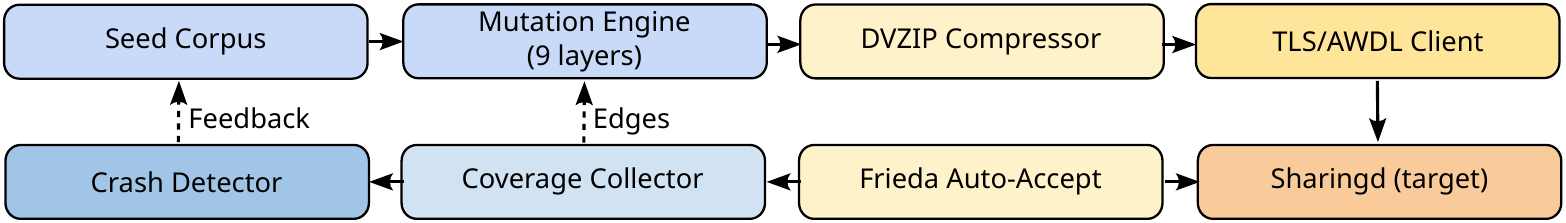}
\caption{\textsc{AirFuzz} architecture. Mutations are applied to seed inputs \emph{before} DVZip compression; Frida hooks bypass the user-acceptance prompt; coverage feedback guides mutation selection.}
\label{fig:airfuzz-arch}
\end{figure*}

We developed \textsc{AirFuzz}, a 12{,}300-line Python protocol-aware fuzzer. Figure~\ref{fig:airfuzz-arch} shows its architecture.

\paragraph{Frida Auto-Accept Daemon.}
To remove the user-interaction dependency, we deploy a Frida-based instrumentation agent on the target machine. The agent hooks five functions in \texttt{sharingd} three ObjC auto-accept predicates, the native \texttt{canAutoAccept} at offset \texttt{+0x041f10}, and the \texttt{UNUserNotificationCenter} notification suppressor, forcing acceptance while leaving all other code paths unmodified. This enables fully automated fuzzing without user interaction. The daemon monitors the target process and automatically reattaches after crashes with configurable recovery using \texttt{launchctl} bootout/bootstrap to reset launchd throttling.

\paragraph{Mutation Engine.}
\textsc{AirFuzz} implements nine mutation layers spanning the protocol stack: HTTP (24 strategies: path corruption, header smuggling, content-length mismatch, chunked-encoding abuse, method fuzzing), binary plist (12 strategies: type confusion, key deletion, oversized strings, nested-container injection, cross-type substitution), DVZip (chunk-length corruption, truncated/oversized chunks, adaptive-mode confusion), CPIO (header corruption, path traversal, symlink/device-node injection, deep path nesting), DER (ASN.1 length overflows, zero-key injection), havoc (random bit flips and byte insert/delete), memcorrupt (25 strategies: boundary values, arithmetic, dictionary ops, block shuffling), and field-level mutations that target specific protocol keys while preserving overall structure.

\paragraph{Pre-Compression Mutation (Key Insight).}
Our initial fuzzer (AF1--AF3) mutated the \emph{compressed} DVZip output, resulting in near-zero (${<}1\%$) server acceptance because even single-byte mutations in the compressed stream corrupt the zlib checksum. The critical design insight in v4 was to mutate the raw CPIO archive \emph{before} DVZip compression. This ensures that the compressed output has valid zlib framing while carrying mutated content, increasing the acceptance rate to over 90\% and enabling meaningful exploration of the archive extraction code paths.

\subsection{Fuzzer Evolution}
\label{sec:fuzz-evolution}

\textsc{AirFuzz} evolved through four versions (AF1--AF4) totalling 1.8K$\rightarrow$12.3K~LoC. AF1 (HTTP + plist only) and AF2 (added DVZip/CPIO mutations) had ${<}5\%$ server acceptance but AF2 still surfaced V1 within the first 200 test cases. AF3 added DER/havoc mutations, Frida auto-accept, and coverage-guided feedback, discovering V2 and V3. AF4's key change was \emph{pre-compression} mutation plus memcorrupt and field-level strategies, raising acceptance above 90\% and expanding edge coverage from 120 to 950 edges into \texttt{Foundation} and \texttt{Network.framework}.

\subsection{Seed Corpus and Coverage}
\label{sec:seed-corpus}

We constructed the seed corpus by capturing legitimate AirDrop transfers between two Apple devices, covering six scenarios: single image, multi-file, large file (500\,MB), URL sharing, contact card (VCF), and directory transfer. Using Wireshark with TLS key logging, we extracted 162 seed templates in total: 70 for Discover, 45 Ask, plus 42 complete Upload payloads. We additionally generated synthetic seeds with boundary values and maximum-length strings for 40+ protocol commands. Coverage is collected via Frida Interceptor hooks on $\sim$100 manually selected functions across \texttt{sharingd}, \texttt{Sharing.framework}, and system frameworks, achieving approximately 950 unique edges across 250K+ executions.

\paragraph{Choice of coverage instrumentation.}
Frida's \texttt{Stalker} engine in principle yields basic-block and edge coverage without manual hook selection, and \textsc{fpicker}~\cite{fpicker} (an extension of \textsc{ToothPicker}~\cite{toothpicker}) wraps it for AFL++-style feedback. We initially attempted this route and found that on our test hosts (\textbf{macOS~26.3 / Apple Silicon ARM64e} and \textbf{iOS~26.3 / arm64e}) \texttt{Stalker} reliably crashes the \emph{target} once it begins instrumenting an arm64e thread that executes inside system frameworks: the target is killed with \texttt{SIGKILL} by the kernel, consistent with arm64e codesigning / pointer-authentication enforcement against \texttt{Stalker}'s in-process code rewriting. The Frida agent itself loads and runs fine, baseline \texttt{Interceptor} hooks are stable, and \texttt{Stalker.follow()} on Frida's own injected thread also works; the failure is specific to following arm64e threads that re-enter signed system code, which is exactly where \texttt{sharingd} spends most of its time. We reproduced this in isolation with \texttt{Stalker.follow()} on the main thread of \texttt{/usr/bin/osascript} on macOS~26.3, which kills the target on script load. For the same reason, \textsc{fpicker}'s LibAFL/Rust mode hits identical mitigations and \textsc{TinyInst}/\textsc{Jackalope}~\cite{tinyinst,jackalope} (which performs out-of-process binary rewriting and was used by Project~Zero against \texttt{coreaudiod}~\cite{p0-coreaudiod}) would require porting to AWDL-bound \texttt{sharingd} and re-implementing the seven-layer AirDrop encoding stack as a TinyInst harness, which we leave to future work. We therefore fall back to a smaller, hand-picked set of \texttt{Interceptor} hooks placed at the input boundary of each protocol layer (HTTP path router, plist scanner, DVZip header, CPIO header, DER/SecureMessage parser). This loses fine-grained edge coverage but is stable on arm64e, survives launchd respawns, and is sufficient to discover V1--V3 and to guide AF4's pre-compression mutation strategy.

\subsection{Crash Detection and Recovery}
\label{sec:crash-recovery}

Crash detection is performed via SSH-based health monitoring of the target machine. When \texttt{sharingd} crashes, \texttt{launchd} restarts it automatically. Our fuzzer detects the restart (new PID) and automatically:
\begin{enumerate}[leftmargin=*,topsep=2pt,itemsep=1pt]
  \item Reattaches the Frida auto-accept daemon to the new process.
  \item Copies the triggering input to a crash corpus.
  \item If \texttt{launchd} throttles restarts (consecutive crash count $\geq 4$, resulting in exponentially increasing delays), uses \texttt{launchctl bootout}/\texttt{bootstrap} to reset the throttle state.
\end{enumerate}

\noindent Crash reports (\texttt{.ips} files) are automatically collected from the target's \texttt{DiagnosticReports} directory for post-mortem analysis.

\section{AirDrop Vulnerabilities}
\label{sec:vulns-airdrop}

We discovered two zero-click and one post-accept pre-authentication vulnerabilities in AirDrop, summarised alongside the Quick Share findings in Table~\ref{tab:vuln-overview} (Section~\ref{sec:introduction}). We detail each AirDrop vulnerability below.

\subsection{V1: Remote DoS via Unhandled HTTP Path}
\label{sec:vuln-v1}

\paragraph{Root Cause.}
The HTTP path router in \texttt{Sharing.framework} (see \path{Network+SFAirDropMessage.swift} in Apple's source layout) maps incoming request paths to \texttt{SFAirDropMessagePath} enum values via a Swift \texttt{switch} statement. Any path that does not match a defined enum value falls into the \texttt{default} case, which calls \texttt{fatalError("Unhandled Path \textbackslash(path)")}. Unlike Swift's \texttt{assertionFailure} (compiled out in release builds), \texttt{fatalError} is unconditionally fatal, so the process traps and is terminated. The vulnerability is therefore a simple, architecture-independent crash: any unauthenticated client that POSTs to an unrecognised URI path forces \texttt{sharingd} to abort. The impact is bounded to a denial of service (no attacker-controlled register or memory state at the crash point) and does not require deep protocol knowledge to reach.

\paragraph{Trigger.}
A single HTTP POST request with any unrecognized URI path and a non-empty body, sent over TLS to port~8770 on the AWDL interface (e.g., \texttt{POST /X HTTP/1.1\textbackslash r\textbackslash nContent-Length:~1\textbackslash r\textbackslash n\textbackslash r\textbackslash nA}).

\paragraph{Impact.}
The crash kills \texttt{sharingd}, immediately disabling AirDrop, AirPlay, Handoff, Universal Clipboard, and Continuity Camera. While \texttt{launchd} restarts the daemon, an attacker can send crash packets in a loop (approximately one every two seconds) to create a \emph{persistent} denial of service. During active loop attack, 0 out of 6 legitimate AirDrop connection attempts succeeded, all 10 attempts succeeded after stopping the attack.

\paragraph{Affected Versions.}
Confirmed on macOS~15.7.3, iOS~18.x, macOS~26.3, and iOS~26.3. The vulnerability is \emph{not present} on iOS~16.7.11, which returns HTTP~400 for unknown paths, suggesting the \texttt{fatalError} was introduced during a Swift port of the path router.

\subsection{V2: Foundation XML Plist Stack Overflow}
\label{sec:vuln-v2}

\paragraph{Root Cause.}
The  \texttt{XMLPlistScanner.scanDict()} of \texttt{Foundation.framework} uses recursive descent parsing with \emph{no depth limit}. Each recursion level consumes approximately 2{,}960 bytes of stack (0x60 bytes for register saves plus 0xb30 bytes for locals). On cooperative thread pools (stack size $\sim$512\,KB), approximately 180--200 levels of nesting exhaust the stack. The next recursive call attempts to write its prologue (saved registers and locals) into the unmapped \emph{stack guard page}, raising a memory \emph{write} fault that the kernel reports as SIGBUS / \texttt{KERN\_PROTECTION\_FAILURE}. The fault address is fully determined by the recursion (the bottom of the cooperative-pool stack) and is not attacker-controlled; no useful register or memory write primitive is exposed.

\paragraph{Trigger.}
An HTTP POST to \texttt{/Discover} with a $\sim$6\,KB XML property list containing $\sim$200 nested \texttt{<dict>} elements (a \texttt{<dict><key>a</key>...} cascade terminated by matching close tags).

\paragraph{Impact.}
This vulnerability affects not just \texttt{sharingd}, but \texttt{Foundation.framework} itself. \emph{Any Apple application} that deserializes untrusted XML property lists via \texttt{PropertyListDecoder} or \texttt{PropertyListSerialization} is vulnerable. The attack surface spans macOS, iOS, watchOS, tvOS, and visionOS. We reproduced the crash on all three architectures Apple currently ships AirDrop on: \textbf{Apple Silicon Mac} (macOS~26.3, ARM64e), \textbf{Intel Mac} (macOS~15.7.4, x86\_64), and \textbf{iPhone with PAC} (iOS~26.3, arm64e). The crash threshold varies slightly with cooperative-thread-pool stack allocation (approximately 200 levels on macOS and 180 on iOS); PAC on arm64e does not change the outcome because the fault originates from the recursive prologue itself rather than a corrupted return pointer. Symbolicated crash logs and reproduction artifacts for each platform are included in the artifact under \path{crash-notes/report-3-sigbus-xmlplist-recursion/}.

\subsection{V3: Network.framework NULL Pointer Dereference}
\label{sec:vuln-v3}

\paragraph{Root Cause.}
Within \texttt{Network.framework}, the HTTP/1.1 connection-setup path \texttt{nw\_protocol\_http1\_connect()} contains a NULL-pointer dereference reachable when the framer is forced into an inconsistent state. The crash happens at a fixed offset into a connection structure whose backing object is \texttt{nil}, so the faulting address sits in the unmapped low page; the precise struct-member offset shifts between builds but the bug is identical across them.

\paragraph{Trigger.}
Crafted HTTP requests with one of: (a) \texttt{Transfer-Encoding: chunked} with negative chunk sizes (\texttt{-1}), (b) duplicate \texttt{Content-Length} headers with conflicting values, or (c) \texttt{Content-Length} vastly exceeding the actual body size.

\paragraph{Impact.}
SIGSEGV (\texttt{KERN\_INVALID\_ADDRESS}) on standard systems where the zero page is unmapped. This vulnerability is in \texttt{Network.framework}'s HTTP/1.1 parser, potentially affecting other Apple applications that process untrusted HTTP traffic through the system networking stack, though the specific trigger conditions (conflicting framing under concurrent load) may limit practical reachability beyond \texttt{sharingd}.

% ============================================================================
% 7. QUICK SHARE VULNERABILITIES
% ============================================================================
\section{Quick Share Vulnerabilities}
\label{sec:vulns-quickshare}

In addition to AirDrop, we analyzed Quick Share implementations across platforms. On a Samsung Galaxy~S23 Ultra running Android~16 (firmware S918BXXS8EZA1, Quick Share v13.8.01.11, GMS~26.05.34), we discovered two vulnerabilities (V4, V5) in the Nearby Connections protocol layer underlying Samsung's Quick Share service. We additionally analyzed Google Quick Share for Windows (version~1.0.2472.1), discovering a critical use-after-free (v6) in the endpoint management subsystem.

\subsection{Quick Share Protocol Analysis}
\label{sec:qs-protocol}

To understand the attack surface, we reverse engineered the Quick Share on Samsung Android by combining protobuf definition extraction from the GMS APK, traffic captures between two Samsung devices, and systematic probing of the TCP listener.

\paragraph{Wire Format.}
Quick Share communicates over TCP using a simple framing protocol: each message is preceded by a 4-byte big-endian length prefix, followed by a Protocol Buffer payload. Application-level messages use the \texttt{OfflineFrame} protobuf, which contains a \texttt{V1Frame} with a \texttt{type} enum field that determines the message category. We identified seven frame types in active use: \texttt{CONNECTION\_REQUEST}~(1), \texttt{CONNECTION\_RESPONSE}~(2), \texttt{PAYLOAD\_TRANSFER}~(3), \texttt{BANDWIDTH\_UPGRADE}~(4), \texttt{KEEP\_ALIVE}~(5), \texttt{DISCONNECTION}~(6), and \texttt{PAIRED\_KEY\_ENCRYPTION}~(7). A \texttt{PAIRED\_KEY\_RESULT}~(8) type is also defined in the binary.

\paragraph{UKEY2 Handshake.}
After the initial \texttt{CONNECTION\_REQUEST}/\texttt{CONNECTION\_RESPONSE} exchange registers the endpoint, UKEY2~\cite{yair2024quickshell} performs mutual authentication via an Elliptic-Curve Diffie-Hellman (ECDH) key agreement on NIST~P-256. The handshake proceeds in three messages using the same 4-byte length-prefix framing but with raw \texttt{Ukey2Message} protobufs (not wrapped in \texttt{OfflineFrame}): \emph{ClientInit} (client $\rightarrow$ server, containing a SHA-512 commitment hash of the \emph{ClientFinished} message, a random nonce, and the selected cipher suite), \emph{ServerInit} (server $\rightarrow$ client, containing the server's ephemeral P-256 public key, a random nonce, and the selected cipher suite), and \emph{ClientFinished} (client $\rightarrow$ server, revealing the client's ephemeral P-256 public key). This commit-then-reveal scheme prevents an attacker from choosing a public key that is a function of the peer's key. Both parties then derive a shared secret via ECDH and expand it using HKDF-SHA256 with a protocol-specific salt (\texttt{D2D\_SALT}, 32 bytes, hardcoded in GMS) to produce per-direction session keys for AES-256-CBC encryption and HMAC-SHA256 authentication.

\begin{figure}[t]
\centering
\includegraphics[width=0.8\columnwidth]{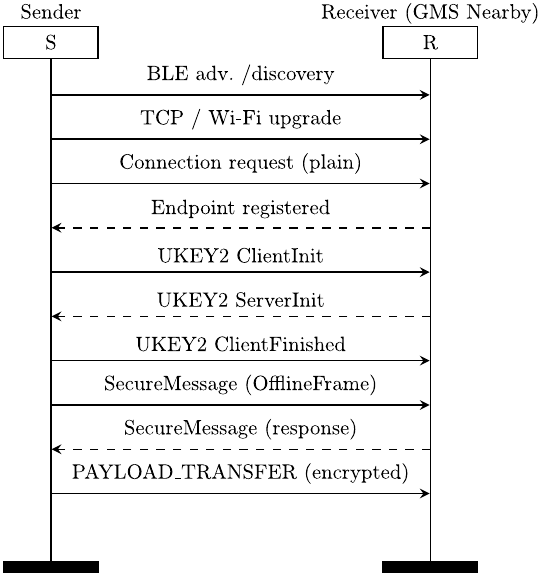}
\caption{Intended Quick Share connection flow. V4 sends \texttt{OfflineFrame}s above the dashed encryption boundary (before UKEY2 finishes), and V5 sends three \texttt{OfflineFrame} types in plaintext below the boundary (after UKEY2 finishes), in both cases bypassing the \texttt{SecureMessage} envelope.}
\label{fig:qs-flow}
\end{figure}

\paragraph{D2D Encryption Layer.}
After UKEY2 completes, all subsequent frames are expected to use the D2D \texttt{SecureMessage} wire format: the plaintext \texttt{OfflineFrame} protobuf is encrypted with AES-256-CBC (IV prepended), then an HMAC-SHA256 tag is computed over the ciphertext. The encrypted payload and tag are wrapped in a \texttt{SecureMessage} protobuf containing \texttt{header\_and\_body} (the encrypted frame) and \texttt{signature} (the HMAC tag). On the receiver side, the \texttt{D2DConnectionContext} class is responsible for unwrapping, verifying, and decrypting \texttt{SecureMessage} envelopes before passing the plaintext frame to the dispatcher. As we show in V4 and V5, this decryption step is not enforced uniformly. We detail the three discovered vulnerabilities below. Figure~\ref{fig:qs-flow} shows the intended Quick Share connection flow that V4/V5 violate.

\subsection{V4: Pre-Authentication Frame Processing Bypass}
\label{sec:vuln-V4}

\paragraph{Root Cause.}
The Nearby Connections protocol used by Samsung Quick Share is designed to enforce a strict handshake sequence: a \texttt{ConnectionRequest} registers the endpoint, then UKEY2 (ClientInit $\rightarrow$ ServerInit $\rightarrow$ ClientFinished) establishes mutual authentication and derives D2D encryption keys. Only after this four-step handshake should the server process application-level OfflineFrame messages. However, the \texttt{PcpManager} in Google Mobile Services (GMS) begins dispatching OfflineFrame messages to handlers \emph{immediately after Step~1} (ConnectionRequest), without waiting for UKEY2 to complete. An unauthenticated attacker needs only to send a single, fixed-format \texttt{ConnectionRequest} protobuf (requiring no secrets), after which the server parses, dispatches, and responds to \texttt{KEEP\_ALIVE}, \texttt{BANDWIDTH\_UPGRADE}, and \texttt{CONNECTION\_RESPONSE} frames, despite never having established an authenticated or encrypted session.

\paragraph{Trigger.}
The attacker opens a TCP connection to the Quick Share port (53601 in our tests), sends a valid \texttt{ConnectionRequest}, and immediately follows it with an OfflineFrame of any of the three affected types. No UKEY2 handshake messages are required. The server responds within milliseconds: for \texttt{KEEP\_ALIVE}, a 58-byte UKEY2 error message (\texttt{"Expected, but did not find ClientInit message type"}) confirming the frame was parsed and dispatched to the UKEY2 state machine; for \texttt{BANDWIDTH\_UPGRADE} and \texttt{CONNECTION\_RESPONSE}, similar error responses. We confirmed this at every stage of the connection lifecycle across 5 independent trials each (100\% reproduction rate).

\paragraph{Systematic Verification.}
To characterize the pre-authentication attack surface, we tested frame processing at each stage of the connection lifecycle (5~trials/stage). Without any prior \texttt{ConnectionRequest} (CR) all frames are silently dropped (negative control, 5/5); immediately after a single CR---with \emph{no} UKEY2 message exchanged---\texttt{KEEP\_ALIVE}, \texttt{BANDWIDTH\_UPGRADE}, and \texttt{CONNECTION\_RESPONSE} are all dispatched and produce server responses (5/5 each); after CR$+$ClientInit, \texttt{KEEP\_ALIVE} is still processed (5/5); after CR$+$ClientInit$+$ServerInit, the server transitions to a state that expects only \texttt{ClientFinished} and drops non-handshake frames (5/5); after the full handshake all frames are accepted (positive control, 5/5). Stages~1--1c are the key finding---three frame types are dispatched to handlers before any UKEY2 authentication.

\paragraph{Impact.}
This vulnerability (CWE-287, CWE-306) breaks the fundamental assumption that no application-level interaction occurs before cryptographic authentication: the attacker drives the Quick Share protocol state machine without any authentication; \texttt{KEEP\_ALIVE} injection prevents connection timeouts; and the dispatcher processes attacker-controlled protobuf content before any integrity check, expanding the pre-authentication surface to frame types designed to be post-authentication only. Combined with V5, this lets a fully unauthenticated attacker inject and have processed all three affected frame types without ever establishing encryption.

\subsection{V5: Device-to-Device Encryption Bypass}
\label{sec:vuln-v8}

\paragraph{Root Cause.}
After the UKEY2 handshake completes and derives AES-256-CBC + HMAC-SHA256 session keys, all subsequent OfflineFrame messages \emph{must} be wrapped in a D2D \texttt{SecureMessage} envelope that provides both confidentiality and integrity. However, the frame dispatcher in the \texttt{EndpointChannelManager} routes incoming frames based on the \texttt{V1Frame.type} field \emph{before} checking whether the frame was received through the D2D decryption pathway. As a result, 3 out of 7 post-handshake OfflineFrame types are processed as raw (unencrypted) protobuf: \texttt{CONNECTION\_RESPONSE} (type~2), \texttt{BANDWIDTH\_UPGRADE} (type~4), and \texttt{KEEP\_ALIVE} (type~5). The remaining 4 types (\texttt{PAYLOAD\_TRANSFER}, \texttt{DISCONNECTION}, \texttt{PAIRED\_KEY\_ENCRYPTION}, \texttt{PAIRED\_KEY\_RESULT}) are correctly rejected when sent without encryption, indicating the encryption check is performed per-handler rather than at the dispatcher level and three handlers simply omit it.

\paragraph{Trigger.}
The attacker completes a full UKEY2 handshake with the target device over TCP (ConnectionRequest $\rightarrow$ ClientInit $\rightarrow$ ServerInit $\rightarrow$ ClientFinished $\rightarrow$ ConnectionResponse), deriving D2D encryption keys. The attacker then sends a \emph{raw} OfflineFrame protobuf not wrapped in the mandatory \texttt{SecureMessage} for any of the three affected types. The server processes the frame and responds:
\begin{itemize}[leftmargin=*,topsep=2pt,itemsep=1pt]
  \item \texttt{CONNECTION\_RESPONSE}: 104-byte response; the server advances its connection state machine to ``accepted,'' as confirmed by logcat: \texttt{"Endpoint~[id] has accepted the connection"}.
  \item \texttt{BANDWIDTH\_UPGRADE}: 29-byte response containing a \texttt{SAFE\_TO\_CLOSE} event that \emph{echoes back} the attacker-supplied IP address and port number, proving the server parsed attacker-controlled data from the unencrypted frame.
  \item \texttt{KEEP\_ALIVE}: 16-byte response with \texttt{ack=true}, demonstrating the KeepAlive handler has no encryption check whatsoever.
\end{itemize}

\begin{figure}[t]
\centering
\includegraphics[width=0.95\linewidth]{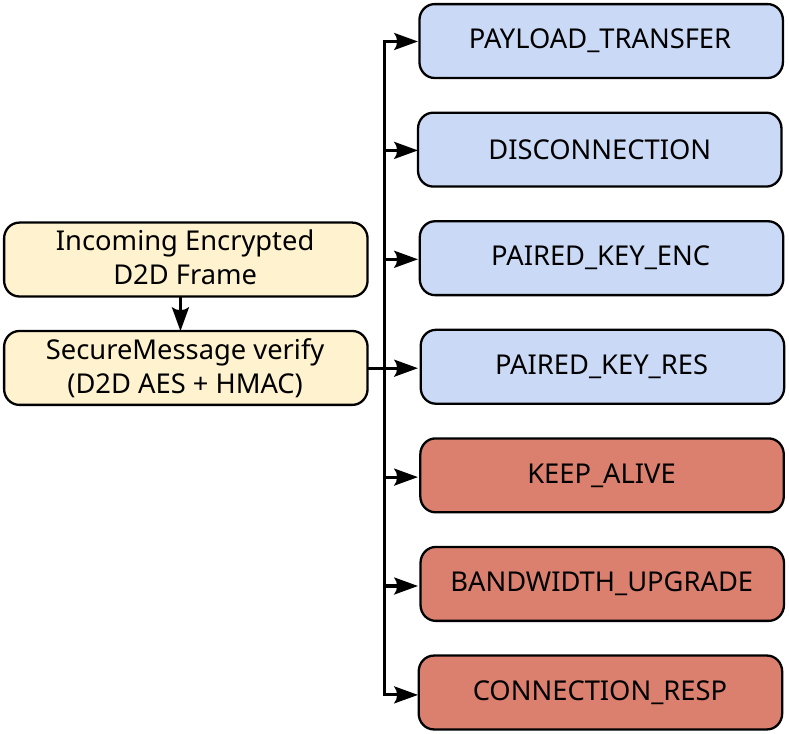}
\caption{V5 per-handler encryption enforcement. Four of seven post-handshake \texttt{OfflineFrame} types are correctly rejected when unencrypted; the three red types are dispatched and processed even when delivered as a raw, unwrapped protobuf, despite the active D2D session.}
\label{fig:v5-enc}
\end{figure}

\noindent We confirmed all three bypasses across 5 independent trials with 100\% reproduction rate on Samsung Galaxy~S23~Ultra running Quick Share~v13.8.01.11 with GMS~26.05.34. Figure~\ref{fig:v5-enc}  highlights which post-handshake frame types cross the \texttt{SecureMessage} boundary and which do not.

%\begin{figure*}[t]
%\centering
%\includegraphics[width=0.7\textwidth]{figs/U4.pdf}
%\caption{V5 per-handler encryption enforcement. Four of seven post-handshake \texttt{OfflineFrame} types are correctly rejected when unencrypted; the three red types are dispatched and processed even when delivered as a raw, unwrapped protobuf, despite the active D2D session.}
%\label{fig:v5-enc}
%\end{figure*}

\paragraph{Impact.}
An on-path attacker on the same \WiFi{} network can inject unencrypted control frames into an active Quick Share session, defeating the D2D encryption layer: unencrypted \texttt{CONNECTION\_RESPONSE} forces a connection into ``accepted'' state, potentially bypassing user consent; \texttt{KEEP\_ALIVE} injection prevents timeouts and keeps sessions alive indefinitely; and unencrypted \texttt{BANDWIDTH\_UPGRADE} frames cause the server to process attacker-supplied IP/port values, leaking endpoint state.

\subsection{v6: Use-After-Free via Endpoint Collision Race Condition}
\label{sec:vuln-v6}

\paragraph{Root Cause.}
Google Quick Share for Windows (version~1.0.2472.1) manages connected peers through dynamically allocated \texttt{EndpointChannel} objects in the Nearby Connections library. We discovered a critical use-after-free (UAF) in the \texttt{OnEncryptionFailed} callback, triggered when \texttt{EncryptionRunner::StartServer()} in \texttt{encryption\_runner.cc} fails the UKEY2 handshake. The vulnerability manifests when two connections arrive with \emph{identical endpoint identifiers and nonce values}: a collision handler in \texttt{base\_pcp\_handler.cc} tears down the endpoint (closing both channels and freeing the \texttt{EndpointChannel} object), while a concurrent thread pool worker executes the encryption failure callback that still holds a raw pointer to the freed object. The stale pointer is subsequently dereferenced for a vtable virtual function call, causing an \texttt{ACCESS\_VIOLATION}.

Notably, the source code at \texttt{base\_pcp\_handler.cc} (line~814) contains a developer comment acknowledging a \emph{previous bug} in this exact code path: \texttt{"We had a bug here, caused by a race with EncryptionRunner."}  The applied fix performs a channel identity comparison (\texttt{*endpoint\_channel != *pending\_connection\_info.channel}), but this comparison \emph{itself dereferences the freed pointer} before comparing, creating a new UAF.

\paragraph{Trigger.}
We developed a multi-phase Python reproducer that rapidly cycles real ECDH handshakes against the Quick Share TCP listener using a \emph{fixed} endpoint identifier (\texttt{"FZ"}), causing register/teardown task queue races. The reproducer proceeds through six phases: sequential handshake cycling (30~rounds), concurrent handshake storm (4~threads $\times$ 15), interleaved collision during handshake (30~rounds), heavy sustained cycling (100~handshakes), final storm (6~threads $\times$ 20), and a combined attack phase. The crash typically occurs when the collision handler logs \texttt{"cleaned up the collision with endpoint FZ by closing both channels. Our nonces were identical''} while a concurrent worker in \texttt{encryption\_runner.cc} attempts to invoke a virtual method on the freed \texttt{EndpointChannel}. The race window is wide ($\sim$11+ seconds of task queue stall), making reproduction reliable.

\paragraph{Crash Analysis.}
Under WinDbg with Page Heap enabled, the crash manifests as a vtable dereference on a freed \texttt{EndpointChannel}: \texttt{mov rcx,[rdi+20h]}; \texttt{mov rax,[rcx]} (read freed vtable); \texttt{call [rax+60h]}. The exception occurs at \texttt{nearby\_share!GetStringFeatureFlagDart+0xaa61e} (file offset \texttt{0x0116fb0e}), with \texttt{rcx} pointing into a PageHeap guard page (\texttt{MEM\_RESERVE} state), confirming the object has been freed. The call stack originates from \texttt{ntdll!TppWorkerThread} through the Nearby Connections thread pool.

\paragraph{Impact.}
The crash primitive is a classic vtable-hijack pattern and exploitation is plausible: (i)~Control Flow Guard is \emph{disabled} in \texttt{nearby\_share.exe}, so no indirect-call target validation is performed; (ii)~without Page Heap, the freed slot falls into the normal process heap and can be reclaimed by a subsequent same-size-class allocation; (iii)~the attacker controls incoming heap-allocated connection data that could reclaim the freed \texttt{EndpointChannel}; (iv)~the wide $\sim$11\,s race window provides ample time for heap spraying between free and use. A standard exploitation path is fake-vtable~$\rightarrow$~ROP~$\rightarrow$~\texttt{VirtualProtect}~$\rightarrow$~shellcode (the usual DEP bypass for CFG-disabled binaries). We confirmed the DoS impact but did not develop a full exploit.

\paragraph{Disclosure.}
We reported V4--V6 through Samsung Mobile Security and the Google Vulnerability Reward Program (VRP). Samsung transferred V4 and V5 to Google after determining that the affected code paths originate in Google's Nearby/Quick~Share components shipped to Samsung; those reports are under investigation at the time of writing. Google's security team acknowledged the validity of V6 against Quick Share for Windows version~1.0.2472.1 and awarded a bounty; a CVE identifier is pending assignment. Section~\ref{sec:ethics} summarises the full disclosure status.

% ============================================================================
% 8. DISCUSSION
% ============================================================================
\section{Discussion}
\label{sec:discussion}

\subsection{Cross-Platform Comparison}

Both protocols expose pre-authentication attack surfaces from wireless proximity (AirDrop's TLS listener accepts any connection; Quick Share's BLE/Wi-Fi-Direct discovery is similarly open) and exhibit insufficient input validation at the application layer---Apple in HTTP path handling and plist parsing, Samsung in pre-authentication frame dispatching and encryption enforcement, Google in endpoint lifecycle management. The designs diverge in encoding (AirDrop's seven-layer HTTP$\rightarrow$bplist$\rightarrow$DVZip$\rightarrow$CPIO stack vs.\ Quick Share's flatter Protocol Buffers) and in concurrency model (AirDrop serializes via a dispatch semaphore, while Quick Share's multi-threaded endpoint manager admits the V6 use-after-free). Each of V1--V6 instantiates a distinct root-cause class---reachable assertion (V1), unbounded recursion (V2), parser state confusion + NULL deref (V3), missing pre-authentication frame check (V4), missing encryption enforcement (V5), and concurrent endpoint UAF (V6)---suggesting that proximity transfer protocols expose a broad spectrum of bug classes rather than a single dominant failure mode. The recurrence of assertion- and parser-class bugs in Apple's code indicates that defensive patterns suited to application reliability (\texttt{fatalError}) become liabilities in network-facing daemons; in Quick Share, V4 and V5 share a single root cause---authentication and encryption checks delegated to individual frame handlers rather than enforced at the dispatcher.

\subsection{Recommended Mitigations}
\label{sec:mitigations}

\paragraph{For Apple AirDrop (V1--V3).} Replace every \texttt{fatalError()} and force-unwrap reachable from network input with \texttt{guard let} and a graceful error return (V1); enforce a maximum nesting depth in \texttt{Foundation}'s \texttt{XMLPlistScanner} (e.g., 64 levels; legitimate AirDrop traffic uses depth~$\leq 4$) together with a maximum HTTP body size (V2); and validate HTTP framing strictly per RFC~9110~\S8.6, rejecting conflicting \texttt{Content-Length} headers, negative chunk sizes, and oversized header values (V3).

\paragraph{For Quick Share (V4--V6).} Reject all \texttt{OfflineFrame} types except UKEY2 handshake messages until authentication completes (V4); decrypt and integrity-verify all post-handshake frames at the dispatcher rather than per-handler (V5); and protect both the endpoint map and endpoint objects with a single lock and defer deallocation via reference counting or epoch-based reclamation until all in-flight operations complete (V6).

\subsection{Broader Impact}
\label{sec:broader-impact}

The vulnerabilities affect a substantial device population: Apple reports over 2.2 billion active devices~\cite{apple-active-devices} running \texttt{sharingd}, and Google reports over 3 billion active Android devices~\cite{google-android-devices} with Quick Share as the default sharing mechanism on Samsung devices and available system-wide on Android. Since V4 and V5 reside in Samsung's Quick Share service layer, further investigation is needed to determine whether other Android OEMs that integrate Quick Share share the same vulnerable code paths. While proximity limits the attack radius to $\sim$10-30\,m, high-density environments (airports, transit, conferences) enable a single attacker to reach hundreds of devices simultaneously.

\subsection{Future Work}
\label{sec:future-work}

Promising directions include escalating V3's NULL-pointer dereference to code execution; verifying V4 and V5 on other Android OEMs that integrate Quick Share,  exploring LLM-assisted decompilation to reduce the manual binary analysis effort our AirDrop reverse engineering required, fuzzing XNU's \texttt{IO80211AWDLPeer} AWDL driver at the link layer, and formally verifying the state machine in Figure~\ref{fig:state-machine} with ProVerif or Tamarin.

\subsection{Ethical Considerations and Responsible Disclosure}
\label{sec:ethics}

All testing was performed exclusively on devices we own in an isolated network environment with no bystander devices in wireless range. All six vulnerabilities were reported to the respective vendors through their coordinated disclosure channels:

\begin{itemize}[leftmargin=*,topsep=2pt,itemsep=1pt]
    \item \textbf{Apple (V1--V3).} Reported to Apple Product Security. All three issues were \emph{acknowledged} and are currently \emph{under fix}; no CVE identifiers have been assigned at the time of writing.
    \item \textbf{Samsung (V4--V5).} Reported to Samsung Mobile Security, who \emph{transferred} the cases to Google after determining that the affected code paths originate in Google's Nearby/Quick~Share components shipped to Samsung.
    \item \textbf{Google (V4--V6).} Reported via Google's Vulnerability Reward Program. V6 (the Windows Quick~Share use-after-free) was acknowledged and Google awarded a bounty for this report; a CVE identifier for V6 is pending assignment. V4 and V5 remain under investigation.
\end{itemize}

\noindent We released AIRFUZZ, crash
reproduction scripts, the Frida auto-accept daemon, protocol documentation, and the seed corpus for WOOT artifact
evaluation on Zenodo: \url{https://zenodo.org/records/20442029}

\subsection{Limitations}

Our analysis has several limitations: \texttt{sharingd} is stripped, so our symbolication relies on export tables, string references, and function boundary heuristics; Frida-based coverage instrumentation may mask or trigger additional race conditions; our Quick Share analysis focused on Samsung Android and Google Windows, other OEMs may differ, and we achieved code execution only once for V6, though V3 to V5 warrant further investigation.

\section{Related Work}
\label{sec:related}

\paragraph{AWDL and AirDrop Protocol Analysis.}
Stute et al.~\cite{stute2018billion,stute2019openwifi,stute2021disrupting} reverse engineered the AWDL link layer, released the open-source OWL implementation, and demonstrated denial-of-service and man-in-the-middle attacks at the wireless layer. Their work forms the substrate on which the present paper's application-layer analysis builds: where OWL exposed AWDL itself, we focus on the HTTP/binary-plist/DVZip/CPIO stack served over AWDL by \texttt{sharingd}. Project~Zero's Beer~\cite{beer2020awdl} subsequently demonstrated a full \iOS{} zero-click radio-proximity exploit against the AWDL kernel driver, showing that the AWDL surface extends below user space; our V1-V3 add three new pre-authentication bugs on the same wireless reach, but in the user-space sharing daemon and its supporting frameworks.

\paragraph{AirDrop Privacy.}
Heinrich et al.~\cite{heinrich2021opendrop} developed OpenDrop, an open-source AirDrop implementation, and analyzed privacy issues in AirDrop's contact discovery and authentication design, showing that exchanged hash-based contact identifiers can enable leakage of phone numbers and email addresses to nearby adversaries. Public reporting has also discussed how such weaknesses may be abused in practice; for example, Bloomberg news ~\cite{bloomberg2024airdrop} reported claims by Chinese authorities about identifying AirDrop message sources. Martin et al.~\cite{martin2019handoff} analyzed privacy properties of Apple's broader Continuity protocol family over BLE, demonstrating identifying and behavioral leakage in related always-on proximity services.

\paragraph{PrivateDrop.}
Heinrich et al.~\cite{heinrich2021privatedrop} proposed PrivateDrop, a privacy-preserving mutual authentication design based on Private Set Intersection (PSI) to replace AirDrop's hash-based contact verification. Apple later introduced an ``Everyone'' receive-mode time limit (first deployed in China with iOS~16.1.1 and then rolled out more broadly) which mitigates some abuse scenarios but does not adopt PSI-style mutual verification~\cite{heinrich2021privatedrop}.

\paragraph{Nearby Connections, Quick Share, and Nearby Share.}
Antonioli et al.~\cite{antonioli2019nearby} reverse engineered and analyzed Google's \emph{Nearby Connections} framework on Android, uncovering design and implementation weaknesses in proximity-based device-to-device communication. More recently, Yair and Cohen~\cite{yair2024quickshell} analyzed Google's Quick Share for Windows and reported a remote code execution attack chain (including CVE-2024-38271 and CVE-2024-38272) exploiting file-handling and protocol-frame processing. Complementing these vulnerability-centric analyses, Kleidermacher~\cite{kleidermacher2025quickshare} describes Google's security-driven design considerations for enabling Android Quick Share interoperability with Apple's AirDrop. Our work differs in focus by examining Samsung's extended Android Quick Share implementation and by developing protocol-aware fuzzing tailored to multilayer proximity-transfer stacks.

\paragraph{Mobile Daemon Fuzzing and Zero-Click Surfaces.}
Protocol-aware fuzzing of mobile system components and always-on daemons has been explored in several contexts. Gro\ss{}~\cite{gross2020remote} demonstrated remote exploitation of the iPhone via iMessage, highlighting the risk of zero-click attack surfaces in background services. Silvanovich~\cite{silvanovich2019zeroclick} systematically analysed the fully remote attack surface of the iPhone, cataloguing zero-click entry points across system services. Closest to our fuzzing methodology, \textsc{ToothPicker}~\cite{toothpicker} fuzzed the \iOS{} Bluetooth daemon over the BLE link layer using Frida-based in-process feedback, and \textsc{fpicker}~\cite{fpicker} extends that approach with AFL++-style mutation, Frida \texttt{Stalker} coverage, and richer mode support; both target the same class of always-on Apple daemons we analyse, and we discuss why their Stalker-based coverage fails on current arm64e \macOS{}/\iOS{} (Section~\ref{sec:fuzzing}). Project~Zero's binary-instrumentation toolchain, \textsc{TinyInst} and \textsc{Jackalope}~\cite{tinyinst,jackalope}, was used to fuzz \texttt{coreaudiod} on \macOS{}~15~\cite{p0-coreaudiod}; porting that out-of-process rewriting model to AWDL-attached \texttt{sharingd} is left to future work. Our fuzzer therefore borrows the protocol-aware, Frida-based posture of ToothPicker/fpicker while replacing Stalker coverage with hand-selected \texttt{Interceptor} hooks at each layer boundary of the AirDrop stack.

\section{Conclusion}
\label{sec:conclusion}

We presented the first cross-platform security analysis of Apple AirDrop 
and Android Quick Share, the two dominant proximity file-transfer protocols 
serving over five billion devices. By reverse engineering AirDrop's 
seven-layer protocol stack-including the undocumented DVZip compression 
format and developing the \textsc{AirFuzz} protocol-aware fuzzer, we 
discovered three pre-authentication vulnerabilities in Apple AirDrop 
(V1--V3): two zero-click and one post-accept. Through complementary manual 
security assessment of Android Quick Share, we found three additional 
vulnerabilities: a pre-authentication frame-processing bypass (V4) and a 
D2D encryption bypass (V5) in Samsung Quick Share, and a use-after-free in 
Google Quick Share for Windows (V6). Our findings show that proximity 
protocols are a \emph{structural} vulnerability class: fatal assertions in 
network-facing code, missing dispatcher-level authentication and encryption 
enforcement, and unsynchronized concurrent endpoint management recur across 
independently developed implementations. All six vulnerabilities have been 
responsibly disclosed; \textsc{AirFuzz} and all reproduction artifacts  
 released as open source.

%%
%% The next two lines define the bibliography style to be used, and
%% the bibliography file.
\bibliographystyle{plain}
\bibliography{references}

%%
%% If your work has an appendix, this is the place to put it.
\appendix

\end{document}